\documentclass{article}
\usepackage{ijcai23}

\usepackage{times}
\usepackage{soul}
\usepackage{url}
\usepackage[hidelinks]{hyperref}
\usepackage[utf8]{inputenc}
\usepackage[small]{caption}
\usepackage{graphicx}
\usepackage{amsmath}
\usepackage{amsthm}
\usepackage{booktabs}
\usepackage{algorithm}
\usepackage{algorithmic}
\usepackage[switch]{lineno}
\usepackage{enumitem}
\usepackage{multirow}
\usepackage{subfigure}
\usepackage{amsfonts}
\usepackage[normalem]{ulem}

\title{Intent-aware Recommendation via Disentangled Graph Contrastive Learning}

\author{
Yuling Wang$^{1}$\and
Xiao Wang$^2$\and
Xiangzhou Huang$^3$\and
Yanhua Yu$^{1}$\footnote{Corresponding author.}\and
Haoyang Li$^4$\and \\
Mengdi Zhang$^3$\and
Zirui Guo$^1$
\And Wei Wu $^3$ \\
\affiliations
$^1$Beijing University of Posts and Telecommunications \\
$^2$Beihang University \\
$^3$Meituan\\
$^4$Tsinghua University
\emails
wangyl0612@bupt.edu.cn, 
xiao\_wang@buaa.edu.cn,
huangxiangzhou@meituan.com,
yuyanhua@bupt.edu.cn,
lihy18@mails.tsinghua.edu.cn,
zhangmengdi02@meituan.com,
zrguo.bupt@qq.com,
wuwei19850318@gmail.com
}

\begin{document}

\maketitle

\begin{abstract}

Graph neural network (GNN) based recommender systems have become one of the mainstream trends due to the powerful learning ability from user behavior data. Understanding the user intents from behavior data is the key to recommender systems, which poses two basic requirements for GNN-based recommender systems. One is how to learn complex and diverse intents especially when the user behavior is usually inadequate in reality. The other is  different behaviors have different intent distributions, so how to establish their relations for a more explainable  recommender system. In this paper, we present the Intent-aware Recommendation via Disentangled Graph Contrastive Learning (IDCL), which simultaneously learns interpretable intents and behavior distributions over those intents. Specifically, we first model the user behavior data as a user-item-concept graph, and design a GNN based behavior disentangling module to learn the different intents. Then we propose the intent-wise contrastive learning to enhance the intent disentangling and meanwhile infer the behavior distributions. Finally, the coding rate reduction regularization is introduced to make the behaviors of different intents orthogonal. Extensive experiments demonstrate the effectiveness of IDCL in terms of substantial improvement and the interpretability.

\end{abstract}

\section{Introduction}
Recommender system provides an effective way to discover user interests and  alleviates the information overload problem.
Recently, recommender systems based on graph neural network (GNN) have attracted much attention, which are able to explore multi-hop relationships of the structural behavior data for better representation~\cite{XiangWang2019NeuralGC,ZihanLin2022ImprovingGC} .
Benefiting from the message passing mechanism of GNN, these graph-based recommender systems are able to  iteratively aggregate the information from neighbors and update user/item nodes, then the high-quality embeddings for user/item can be obtained~\cite{wang2017community,li2024dynamic}.

Despite that traditional GNN based recommender systems are able to fully take advantage of high-order relationships and learn effective user/item representations, most of them are not specifically designed to understand the underlying user intents from behavior data. It is well established that characterizing the complex relationships between observed behaviors and the underlying user intents is the key to recommender systems, which consequently poses two basic requirements for GNN based recommendations. 

\begin{figure}
\centering
  \includegraphics[width= 5cm, height = 2.6cm]{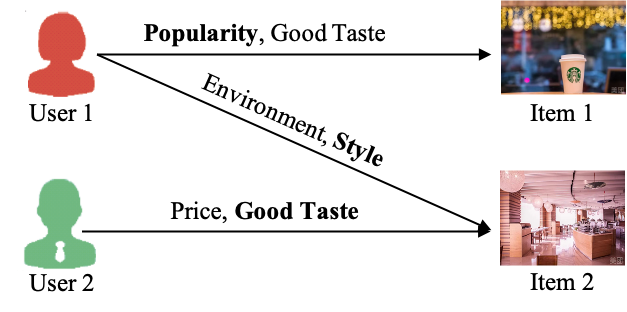}
  \caption{An illustration of behavior distributions. The intents for each behavior are shown on the arrow (the main intents are bolded).}
  \label{example}
\end{figure}

One is \emph{how to effectively learn the complex and diverse intents, especially when the observed behavior data is very sparse in reality?} As shown in Figure~\ref{example}, despite that the user-item behavior can be uniformly represented by an edge in this graph, the underlying user intents actually are very different. 
For example, 
user 1 purchases item 1 due to its popularity and good taste, but the reason she interacts with item 2 is that the environment and style meet her preferences. Meanwhile, 
user 2  exhibits the intents of price and taste towards item 2.
Therefore, discovering the user intents  plays an important role on understanding user behavior, and holds great potential on improving and 
 explaining the recommender systems. Most of previous GNN based methods ignore user intents and directly learn the user/item representations~\cite{XiangnanHe2020LightGCNSA,JiancanWu2021SelfsupervisedGL}, while few works attempt to obtain the user intents using disentangled learning~\cite{XiangWang2020DisentangledGC}, however,
it is well known that a  well-disentangled model usually requires rich  inductive biases and supervision~\cite{FrancescoLocatello2018ChallengingCA}.
As the observed interactions are extremely sparse in reality,
it is highly desirable to characterize more supervision signals from data for  a better disentanglement.

The other is \emph{once the user intents are discovered, how to establish the behavior distribution over user intents?}
As shown in Figure~\ref{example}, user 1 pays more attention to popularity than taste towards item 1. Thus, modeling behavior distribution can describe the strength of different user intentions,
which understands user preference more accurately, i.e.,  we can better learn the behavior representation based on the most closely related user intents.
 Moreover, the behaviors originating from different intents should be distributed in different subspaces as much as possible, which enables the learned behavior representation more discriminative across different intents. 
 Few GNN based methods learn behavior distribution by calculating the similarity between user and item in each intent subspace~\cite{XiangWang2020DisentangledGC}, 
 which ignores the semantic characteristics of intent,
causing the learned distribution may deviate from the specific intent.
Besides, the current GNN based methods still cannot guarantee that behavior representations are correctly distributed across different intent subspaces.

In this paper,  we propose the Intent-aware Recommendation via Disentangled Graph Contrastive Learning (IDCL) to simultaneously learn interpretable user intents and behavior distributions over them.
Firstly, we model the user behavior data as a user-item-concept graph, where the concept represents the multi-aspect semantics of item (e.g., movie genre).
Then an augmented graph can be obtained by perturbing the original graph, and we design a behavior disentangling module to learn the disentangled behavior representations from the two graphs. 
Meanwhile, a set of concept-aware semantic bases is obtained by soft clustering from concept embeddings, each of which can be used as explicit guidance to facilitate disentangling meaningful intent.
We then propose an intent-wise contrastive learning to further enforce disentangling  and infer the behavior distribution.
To promote the behaviors of different intents more independent, 
so that the learned behavior representations are more discriminative across different intents,
we introduce the coding rate reduction regularization.
Our key contributions can be summarized as follows:
\begin{itemize}
\item 
We study the problem that how to effectively learn complex and diverse user intents for better and interpretable GNN based recommender systems, especially when the user behaviors are sparse.
\item 
We propose an Intent-aware Recommendation via Disentangled Graph Contrastive Learning (IDCL) model, which is able to fully utilize the concepts and contrastive learning to learn better disentangled user intents, as well as the behavior distributions.
\item 
Extensive experiments are conducted on three datasets, which  demonstrates the effectiveness of our proposed IDCL.  Futher analysis shows that the learned intent representations and behavior distributions are interpretable.
\end{itemize}

\section{Related Work}
\subsection{GNN based Recommender System}
Traditional shallow recommender systems approach recommendation as a representation learning problem~\cite{YehudaKoren2009MatrixFT,rendle2010factorization}, then some neural models are proposed to incorporate the powerful expressive power of MLP~\cite{HuifengGuo2017DeepFMAF,he2017neural}.
Recently, GNN based recommender systems are proposed to capture the higher-order connectivity by organizing interaction data into a graph and applying the powerful message passing mechanism of GNN\cite{XiangWang2019NeuralGC}.
For instance,
LightGCN obtains promising results by simplifying the components of GCN and applying it to  user-item interaction graph~\cite{XiangnanHe2020LightGCNSA}.
Moreover, some studies propose to incorporate self-supervised learning to alleviate the data sparsity problem~\cite{JiancanWu2021SelfsupervisedGL,ZihanLin2022ImprovingGC}.
To better understand the user intents, DGCF uses disentangled learning for GNN-based recommendation~\cite{XiangWang2020DisentangledGC}.
More GNN based recommender systems can refer to~\cite{wu2022graph}.

\subsection{Disentangled Representation Learning}
Disentangled learning is well researched in the field of computer vision, including supervised learning based methods~\cite{ZhenyaoZhu2014MultiViewPA,JunTingHsieh2018LearningTD} and some unsupervised  methods~\cite{XiChen2016InfoGANIR,IrinaHiggins2017betaVAELB}.
Recently, DisenGCN introduces disentangled  learning to graph-structured data to learn disentangled node representation~\cite{JianxinMa2019DisentangledGC}.
DGCL uses contrastive learning to
 identify the latent factors in graph and derives the disentangled graph representation~\cite{HaoyangLi2021DisentangledCL}.
Moreover, disentangled learning also brings new opportunities for  recommendations, which learns fine-grained user interests from observed behaviors to boost both the performance and interpretability~\cite{ma2019learning,wang2020multi,zhang2022geometric}.
For instance, 
~\cite{ma2019learning} achieves both macro disentanglement and micro disentanglement based on a generative model.
~\cite{zhang2022geometric} achieves disentangling across multiple geometric spaces. 
 Additionally, ~\cite{XinWang2022DisentangledRL,guo2022topicvae} introduce additional knowledge to variational autoencoder to guide the meaningful disentangling.

\section{Methodology}
\label{method}
\begin{figure*}
  \includegraphics[width=\textwidth]{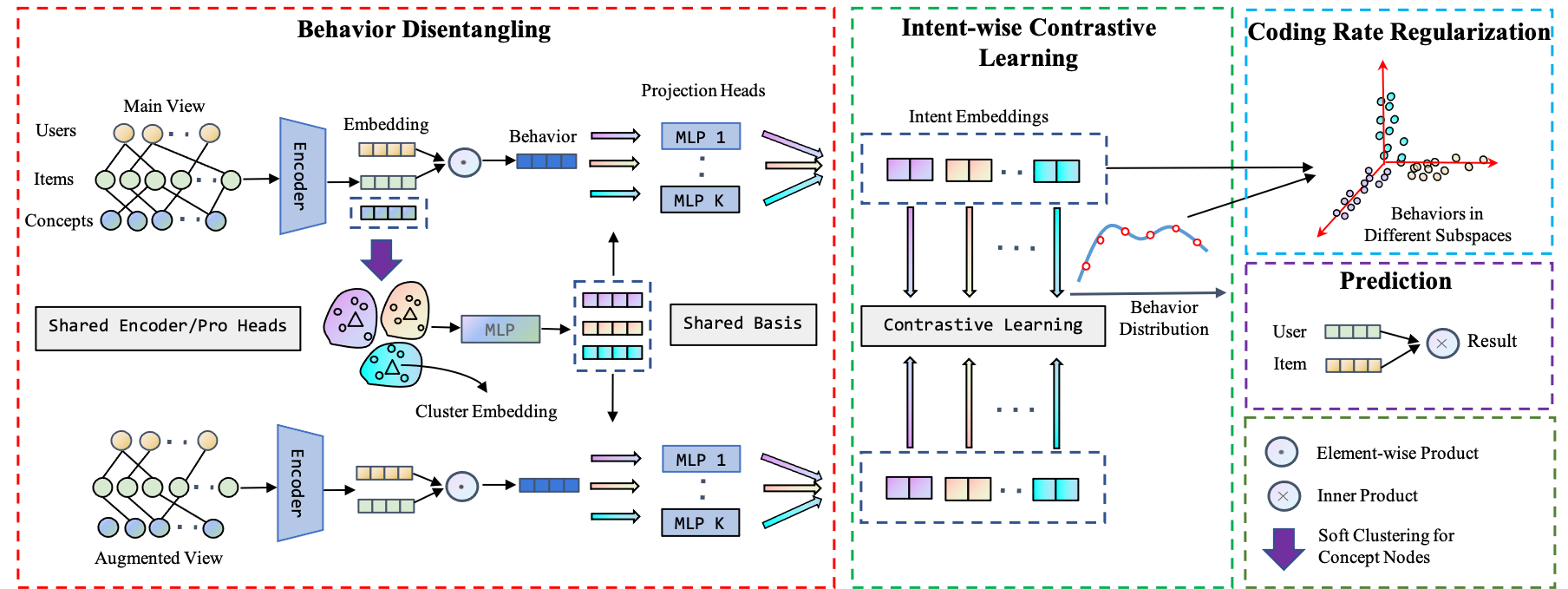}
  \caption{The framework of the proposed Intent-aware Recommendation via Disentangled Graph Contrastive Learning (IDCL)  model.}
  \label{model}
\end{figure*}

In this section, we introduce the proposed IDCL model (Figure~\ref{model}), which mainly contains four modules: Behavior Disentangling (BA), Intent-wise Contrastive Learning (ICL), Coding Rate Reduction Regularization (CR) and Prediction.
The workflow of IDCL is as follows.  
We first model user historical behavior data as a user-item-concept graph, 
and the augmented graph is constructed via edge dropout,
then BA module takes as input the two graphs to discover diverse user intents and infer a set of concept-aware semantic bases.
Then, the ICL module is proposed to enhance the intent disentangling and provide fine-grained self-supervised information, while the  behavior distributions are inferred via a semantic basis based method.
Besides, as an information theory based criterion, the CR module acts as a regularization constrain to promote the orthogonality between behaviors of different intents.
Finally, the model makes the prediction based on the learned  representations of user and item.

\subsection{Problem Definiton}

\paragraph{Multi-intent based prediction.}
Usually, the user behavior data can be typically represented by a graph $\mathcal{G}=(\mathcal{V}, \mathcal{E})$, where the node set $\mathcal{V}=\mathcal{U} \cup \mathcal{I} \cup \mathcal{C}$ involves all users, items and item-related concepts, and the edge set $\mathcal{E}=\mathcal{O}^{+} \cup \mathcal{P}^{+}$ represents the observed user behaviors and item affiliations. Specifically, $\mathcal{U}=\left\{u_1, u_2, \cdots, u_N\right\}$ is the set of $N$ users, $\mathcal{I}=\left\{i_1, i_2, \cdots, i_M\right\}$ is the set of $M$ items, $\mathcal{C}=$ $\left\{c_1, c_2, \cdots, c_R\right\}$ is the the set of $R$ item-related concepts which express the item characteristics, such as category, genre,
popularity, etc.  $\mathcal{O}^{+}=\left\{e_{u i}| u \in \mathcal{U}, i \in \mathcal{I}\right\}$
 represents the $F$ historical behaviors between users and items,
 where $e_{u i}$ indicates that user $u$ has adopted item $i$ before.   $\mathcal{P}^{+}=\left\{b_{i c}| i \in \mathcal{I}, c \in \mathcal{C}\right\}$
  indicates that item $i$ belongs to concept $c$.
 Given a candidate pair $(u, i)$ consisting of a target user $u$ and a potential item $i$, our goal is to learn users' disentangled intents as well as the behavior distributions over intents and then predict $y_{u i} \in\{0,1\}$ , which indicates how likely this item should be recommended to the target user. 

\subsection{Behavior Disentangling}
\label{bd}
A user's adoption of an item could be driven by multiple complex intents, which are usually closely related to user's personality and item characteristics,
ignoring any side of information  may result in the insufficient and inaccurate intent modeling.
Thus, it is desired to disentangle the underlying intents from behavior using the combination of user and item. 


Given a user behavior graph  $\mathcal{G}=(\mathcal{V}, \mathcal{E})$, the widely used LightGCN~\cite{XiangnanHe2020LightGCNSA} acts as a graph encoder to learn the $d$-dimensional representations with high-order collaborative signals of user, item, and concept.
To be specific, the representation $\mathbf{z}_{u_i}^{(l)}\in \mathbb{R}^{d}$ of user $u_i$ at $l$-th layer can be obtained by aggregating the information of neighborhoods based on message passing mechanism of GNN as follows: 
\begin{equation}
    \mathbf{z}_{u_i}^{(l)}=f_{\text {c }}\left(\mathbf{z}_{u_i}^{(l-1)}, f_{\text {a }}\left(\left\{\mathbf{z}_{i_j}^{(l-1)}| i_j \in \mathcal{N}_{u_i}\right\}\right)\right),
\end{equation}
where $f_{\text {a }}(\cdot)$ and $f_{\text {c}}(\cdot)$  are aggregate and combine functions respectively, and they have multiple choices in different GNNs~\cite{KeyuluXu2018RepresentationLO,ThomasKipf2016SemiSupervisedCW}. We then employ a readout function $f_{\text {r }}(\cdot)$  that integrates the representations from different layers to obtain  the final representation:
\begin{equation}
    \mathbf{z}_{u_i}=f_{\text {r }}\left(\left\{\mathbf{z}_{u_i}^{(l)}| l=[0, \cdots, L]\right\}\right).
\end{equation}
Similarly, we can obtain the representation $\mathbf{z}_{i_j}$ of item $i_j$ and the representation $\mathbf{z}_{c_r}$ of concept $c_r$.
Then the  representation $\mathbf{z}_e$ of behavior that user $u_i$ interacts with item $i_j$ is:
\begin{equation}
 \mathbf{z}_e =    \mathbf{z}_{u_i} \odot \mathbf{z}_{i_j}.
\end{equation}

Unlike previous works that perform disentangling on user $\mathbf{z}_{u_i}$ or item $\mathbf{z}_{i_j}$ individually~\cite{JianxinMa2019DisentangledGC,XiangWang2020DisentangledGC}, 
we operate directly on behavior $\mathbf{z}_e$, which combines both of user and item representations.
Assuming that there are $K$ latent intents causing the behaviors,  which associated with item-related concepts to some extent.
Since each concept aggregates the semantics from all items with the same aspect attribute in the graph.
We  extract  $K$ high-level semantic bases from item-related concepts $\mathbf{Z}_{c} = \left\{\mathbf{z}_{c_r}\right \}_{r=1}^R $ via soft clustering~\cite{ying2018hierarchical}.
Firstly, a probabilistic concept assignment matrix is learned as:
\begin{equation}
\mathbf{S} = \text{softmax} \left(\mathbf{Z}_{c}\mathbf{W}_{1}
\right) \in \mathbb{R}^{R\times K},
\end{equation}
where $ \mathbf{W}_{1} \in \mathbb{R}^{d\times K}$,  and $K$ is a model hyperparameter. Each row of  $\mathbf{S}$ provides a soft assignment of the concept node to different intents. 
Then we aggregate concept nodes $\mathbf{Z}_{c}$ according to the assignment $\mathbf{S}$, resulting $K$ cluster embeddings, then a semantic projection head $g_{s}(\cdot)$ performs feature transformation on those cluster embeddings and  outputs a set of concept-aware semantic bases as follows:
\begin{equation}
\label{basis}
\mathbf{Z}_{B} = g_{s}(\mathbf{S}^\top \mathbf{Z}_{c} ) \in \mathbb{R}^{K \times \Delta d},
\end{equation}
where $\Delta d=d / K$, and $\mathbf{Z}_{B}=\left\{\mathbf{b}_{k}\right \}_{k=1}^K$,  each of which corresponds to a different semantic space, then they serve as semantic guidance and are combined with the behavior embedding to facilitate disentangling meaningful intent:
\begin{equation}
\label{intentk}
\mathbf{z}_{e, k}=g_{b}^{(k)}\left(\mathbf{z}_e \mathbin\Vert \mathbf{b}_{k} \right) \in \mathbb{R}^{\Delta d},
\end{equation}
where $ \mathbin\Vert$ means the concatenation of two embeddings,  $g_{b}^{(k)}$ is the behavior projection head that maps the combination to the $k^{th}$ intent spaces, and $\mathbf{z}_{e, k}$ indicates the $k^{th}$ intent.
Analogously, Eq.~\eqref{intentk} is also applied to  calculate all remaining intents via separatly projection heads in $g_{b}(\cdot)=\left\{g_{b}^{(k)}(\cdot)\right\}_{k=1}^K$.
The final disentangled behavior representation can be obtained by combining all intents: $\mathbf{z}_e=\left[\mathbf{z}_{e, 1}; \mathbf{z}_{e, 2}; \ldots; \mathbf{z}_{e, K}\right]$.

\subsection{Intent-wise Contrastive Learning}
As there is usually no intent-wise labeled data in reality, however, disentanglement learning highly desires to consider the role of (implicit) supervision~\cite{FrancescoLocatello2018ChallengingCA}.
Additionally, behavior distributions over intents can further reflect the strength of different user intents, increasing the interpretability of recommendation. Thus, in this module, a intent-wise contrastive learning is designed to enforce meaningful disentangling  and infer  behavior distributions.


We first construct the augmented graph $\mathcal{G}^{\prime}$ for the original graph $\mathcal{G}$ through widely used edge dropout strategy~\cite{JiancanWu2021SelfsupervisedGL}, and the shared graph encoder and behavior projection heads are all applied to the augmented view, then we get the augmented factorized behavior  embedding 
${\mathbf{z}_e}^{\prime} = \left[\mathbf{z}^{\prime}_{e, 1}, \mathbf{z}^{\prime}_{e, 2}, \ldots, \mathbf{z}^{\prime}_{e, K}\right]$.
Thus, as ~\cite{HaoyangLi2021DisentangledCL}, the intent-wise contrastive learning loss can be defined as follows:
\begin{equation}
    \mathcal{L}_{icl}= \sum_{e\in \mathcal{O}^{+}} -log \mathbb{E}_{p_{\theta}(k| e)}[p_{\theta}(e^{\prime}| e,k)],
\end{equation}
where $p_{\theta}(k| e)$ indicates the probability  over the $k^{th}$  intent of  behavior $e$, and $p_{\theta}(e^{\prime}| e,k)$ is the behavior  contrastive learning subtask under the $k^{th}$ intent. We aim to learn the optimal $K$ intents which are able to maximize the expectation of $K$ subtasks.
The behavior confidence over the  $k^{th}$ intent is inferred based on the concept-aware semantic basis $\mathbf{b}_k$  as:
\begin{equation}
    \label{ek}
    p_\theta\left(k| e\right)=\frac{\exp \phi\left(\mathbf{z}_{e, k}, \mathbf{b}_k\right)}{\sum_{k=1}^K \exp \phi\left(\mathbf{z}_{e, k}, \mathbf{b}_k\right)},
\end{equation}
where $\mathbf{b}_k$ is calculated from Eq.~\eqref{basis}, $\phi$ is the cosine similarity with temperature $\tau$, and $\sum_{k=1}^K p_\theta\left(k| e\right) = 1$,
which ensures that intents with high confidence are more likely to have a greater impact on contrastive learning. 
It is worth mentioning that  we utilize $\left\{\mathbf{b}_k \right\}_{k=1}^K$  as  prototypes instead of random initialization~\cite{HaoyangLi2021DisentangledCL,ma2019learning}, which incorporates the interpretable signals from item-related concepts. 

The contrastive learning subtask of the $k^{th}$ intent is:
\begin{equation}
\label{contrast}
    p_\theta(e^{\prime}|e,k)=\frac{\exp \phi\left(\mathbf{z}_{e, k}, \mathbf{z}_{e, k}^{\prime}\right)}{\sum_{j \in \mathcal{O}^{+}, j \neq e}^{|\mathcal{O}^{+}|} \exp \phi \left(\mathbf{z}_{e, k}, \mathbf{z}_{j, k}^{\prime}\right)},
\end{equation}
where $\mathbf{z}_{e, k}$ and $\mathbf{z}_{e, k}^{\prime}$ are the positive pair of the $k^{th}$ intent.
To reduce the computational complexity, we use the NT-Xent loss on a minibatch and randomly sample a portion of behaviors from each training batch~\cite{TingChen2020ASF}.




\subsection{Coding Rate Reduction Regularization }
As behaviors driven by different intents should be distributed in different subspaces, which enables the learned behavior representations more discriminative according to intents.
Here we utilize maximizing coding rate reduction ($\text{MCR}^{2}$)~\cite{yu2020learning} as a geometric regularizer for behavior representations, which measures the volume difference between  representations of the entire behaviors and each intent group of behaviors.
It is worth mentioning that $\text{MCR}^{2}$ considers the intrinsic geometric of features, which is able to enhance the diversity of behavior representations.

Firstly, we compute the coding rate of all behaviors, where a higher coding rate indicates that more space is required to encode the representations.  Given  behavior representations $\mathbf{Z}_{e} = \left\{\mathbf{z}_{e}\right \}_{e=1}^F \in \mathbb{R}^{F \times d}$, the coding rate  for the whole behaviors is defined as 
 the average coding length per behavior
~\cite{ma2007segmentation},  which is formulated as follows:
\begin{equation}
\label{cr}
 R(\mathbf{Z}_{e},\epsilon) =  \frac{1}{2} \log \operatorname{det}\left(\mathbf{I}+\frac{d}{F \epsilon^{2}} {\mathbf{Z}_{e}}^{\top} \mathbf{Z}_{e} \right),
\end{equation}
where $\epsilon$ is a tolerated hyperparameter, which denotes the expected decoding error is less than  $\epsilon$.


In fact, we tend to map the behaviors driven by different intents into different  subspaces, keeping them as orthogonal as possible. 
 Fortunately, Eq.~\eqref{ek} provides a soft assignment of each behavior to $K$ intent groups. We define a set of membership matrices
$\boldsymbol{\Pi}=\left\{\boldsymbol{\Pi}_k \in \mathbb{R}^{F \times F}\right\}_{k=1}^K$,
where $\boldsymbol{\Pi}_k$ is the diagonal matrix whose diagonal element is the  probability of each behavior subject to the $k^{th}$ intent, i.e., $ p_\theta\left(k| e\right)$ in Eq.~\eqref{ek}. 
If each behavior group is coded separately,
the $k^{th}$ group has an expected number of  $\operatorname{tr}\left(\boldsymbol{\Pi}_k\right)$
vectors. 
 Thus, with respect to partition $\boldsymbol{\Pi}$,
 the total compactness for each group of behaviors  is the summation of coding rate for all behavior groups:
\begin{equation}
\resizebox{.91\linewidth}{!}{$
            \displaystyle
    R^c(\mathbf{Z}_{e}, \epsilon | \boldsymbol{\Pi}) \doteq  
    \sum_{k=1}^K \frac{\operatorname{tr}\left(\boldsymbol{\Pi}_k\right)}{2F} \log \operatorname{det}\left(\boldsymbol{I}+\frac{d}{\operatorname{tr}\left(\boldsymbol{\Pi}_k\right) \epsilon^2} {\mathbf{Z}_{e}}^{\top} \boldsymbol{\Pi}_k \mathbf{Z}_{e}\right) $}.
\end{equation}
The volume difference between representations of the whole  and each group of behaviors is desired lager, i.e., maximizing the coding rate reduction brings a better representation:
\begin{equation}
    %
   \mathcal{L}_{\Delta R} =  -R(\mathbf{Z}_{e}, \epsilon) + R^c(\mathbf{Z}_{e}, \epsilon | \boldsymbol{\Pi}),
\end{equation}
where the first term expands the diverse feature space of all behaviors, and the second term enforces more similar representations for behaviors within the same intent group.

\begin{table*} [t]
\centering
 
  \begin{tabular}{clcccc}
    \toprule
    \multirow{2}{*}{Dataset} & \multirow{2}{*}{Method }&  \multicolumn{4}{c}{Metrics} \\
    \cline{3-6}
    &            &Recall@20    &Recall@50    &Recall@100        &NDCG@100\\
    \multirow{6}{*}{ML-100k}
    &NGCF        &0.2395±0.0379 &0.3885±0.0442  &0.5123±0.0454 &0.2758±0.0296  \\
    &LightGCN    &0.2724±0.0175 &0.3878±0.0255  &0.5278±0.0185 & 0.2975±0.0182\\
    &DGCF        &0.2371±0.0369 &0.3847±0.0264  &0.5096±0.0291 & 0.2858±0.0234 \\
    &MacidVAE    &0.2981±0.0384 &0.4287±0.0175  &0.5378±0.0317 & 0.3210±0.0176 \\
    &NCL         &0.2347±0.0191&0.3771±0.0175 &0.5096±0.0291& 0.2796±0.0201 \\
    &IDCL    &\textbf{0.3235±0.0073}&\textbf{0.4450±0.0083}&\textbf{0.5554 ±0.0045}& \textbf{0.3378±0.0078}  \\
    
    \hline
    \multirow{6}{*}{ML-1M}
    &NGCF        & 0.2678±0.0171&0.4294±0.0177&0.5734±0.0221   & 0.3856±0.0148 \\
    &LightGCN    & 0.2940±0.0097&0.4694±0.0194&0.6125±0.0172& 0.4150±0.0117\\
    &DGCF        & 0.2961±0.0050 &0.4664±0.0054&0.6073±0.0018 & 0.4115±0.0015 \\
    &MacidVAE    & 0.2981±0.0060&0.4590±0.0053&0.5988±0.0053&0.4104±0.0045\\
    &NCL         & 0.3017±0.0043&0.4754±0.0055&0.6175±0.0040&0.4177±0.0028\\
    &IDCL        &\textbf{0.3160 ±0.0030}&\textbf{0.4888±0.0030}&\textbf{0.6268±0.0028}&\textbf{0.4302±0.0017} \\
    
    \hline
    
    \multirow{6}{*}{MtBusiness}
    &NGCF        &0.2768±0.0022 & 0.3088±0.0013 & 0.3303±0.0005 &0.2258±0.0015   \\
    &LightGCN    &0.2934±0.0024 & 0.3354±0.0008 & 0.3597±0.0007 &0.2378±0.0023  \\
    &DGCF        &0.2915±0.0024& 0.3318±0.0054  &0.3541±0.0071 & 0.2358±0.0016\\
    &MacidVAE    & 0.2887±0.0013 &0.3309±0.0010 &0.3569±0.0027 & 0.2333±0.0012\\
    &NCL         & 0.2906±0.0021& 0.3353±0.0015 & 0.3605±0.0015 & 0.2335±0.0035\\
    &IDCL        &\textbf{0.2973±0.0010}&\textbf{0.3426±0.0011 }&\textbf{0.3697±0.0014}&\textbf{0.2382±0.0003}\\

    \bottomrule
  \end{tabular}
  
  \caption{The recommendation performance comparison. Best results are in bold.}
   \label{total}
\end{table*}

\subsection{Prediction}
Based on the learned  representations of user and item, 
the preference score of user $u$ towards item $i$ can be predicted as:
\begin{equation}
    \hat{y}_{ui} = \mathbf{z}_{u}^\top \mathbf{z}_{i}.
\end{equation}
We use pairwise Bayesian Personalized Ranking (BPR) loss~\cite{SteffenRendle2009BPRBP}, which promotes higher score for the observed positive pair $(u,i) \in \mathcal{O}^{+}$ than the unobserved counterparts $(u,j) \in \mathcal{O}^{-}$ as follows:
\begin{equation}
    \mathcal{L}_{bpr}=\sum_{(u, i, j) \in O}-\log \sigma\left(\hat{y}_{u i}-\hat{y}_{u j}\right).
\end{equation}

In addition to model user-item interaction, we treat the proposed two losses as supplementary and design a multi-task training loss to jointly 
optimize the traditional  recommendation loss $\mathcal{L}_{bpr}$,   the  self-supervised loss  $\mathcal{L}_{icl}$ and the  coding rate reduction regularization $\mathcal{L}_{\Delta R}$:
\begin{equation}
    \label{loss}
 \mathcal{L}=\mathcal{L}_{bpr}+\lambda_1 \mathcal{L}_{icl}+ \lambda_2 \mathcal{L}_{\Delta R}+\lambda_3\|\Theta\|_2^2,   
\end{equation}
where $\Theta$ is the set of model parameters, $\lambda_1$, $\lambda_2$ and $\lambda_3$ are hyperparameters to control the strengths of each components.
        

\section{Experiment}
\subsection{Experimental Settings}
\noindent
 \paragraph{Datasets.}
We conduct our experiments on three real-world datasets. 
In detail, for two MovieLens datasets with different scales (i.e., ML-100k, ML-1M)~\cite{FMaxwellHarper2015TheMD}, we follow the split method in MultiVAE~\cite{liang2018variational} and MacridVAE~\cite{ma2019learning}, and the movie genres are used as concepts. 
In addition, 
we also collect a dataset from the platform recommender system of Mobile Meituan App\footnote{\url{http://i.meituan.com/}},  named MtBusiness, 
including 52041 purchase records of 11891 users to 20689 businesses, and it involves the multi-aspect business information as concepts that is suitable for disentangled learning. 
We split all users into training/validation/test sets as MultiVAE, then we select 4000 held-out users,  for each held-out user, we randomly choose $50\%$ of the interactions to report metrics. 


%
\noindent
 \paragraph{Baselines.}
We compare IDCL with five SOTA baselines. Among them, NGCF~\cite{XiangWang2019NeuralGC}, LightGCN~\cite{XiangnanHe2020LightGCNSA} and  NCL~\cite{ZihanLin2022ImprovingGC}  as three popular GNN-based recommendation approaches are included. In addition, we include two recently proposed disentangled  recommendation models, i.e.,  MacridVAE~\cite{ma2019learning}    and DGCF~\cite{XiangWang2020DisentangledGC}.
\noindent
 \paragraph{Evaluation Metrics.}
Following~\cite{XiangWang2019NeuralGC}, for users in the testing set, we use the all-ranking protocol to evaluate the top-$K$ recommendation performance. We adopt two popular metrics for evaluation: Recall$@K$ and NDCG$@K$, where $K\in\{20, 50, 100\}$, and we report the average scores of 5 runs and standard deviation.
\noindent
 \paragraph{Implementation and hyper-parameters.}
We implement our model based on Pytorch.\footnote{\url{https://pytorch.org/}}
We conduct experiments  on all datasets with the fixed training/validation/test split. 
We implement all the baselines with the unified opensource of recommendation algorithms, i.e.,  RecBole ~\footnote{https://github.com/RUCAIBox/RecBole}~\cite{WayneXinZhao2020RecBoleTA}.
To make a fair and reliable comparison, we take the same item-related concept information as the initial feature for all baselines, and we carefully search hyper-parameters of all the baselines  to get the best performance.
We keep the embedding size of ours and all baselines to be the same, and the GNN layers of ours and all GNN-based baselines are set to be consistent.
We employ the early stopping strategy for all experiments to prevent overfitting. 
The Adam optimizerfor mini-batch gradient descent is applied to train all models.
We turn the hyper-parameters in validation set using random search, and the search space of some important hyper-parameters are:
$K \in  \{6 , 8, 10, 12, 14, 16\}$,  $\Delta d \in [20, 40]$.

\begin{figure*}[t]
\centering
\subfigure[Behavior]{
\label{heat1}
\includegraphics[width=4cm,height = 2.8cm]{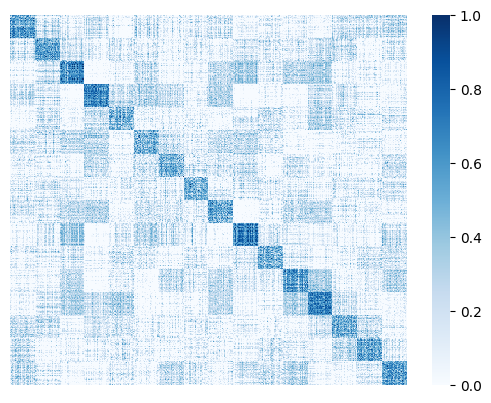}}
\subfigure[User]{
\label{heat2}
\includegraphics[width=4cm,height = 2.8cm]
{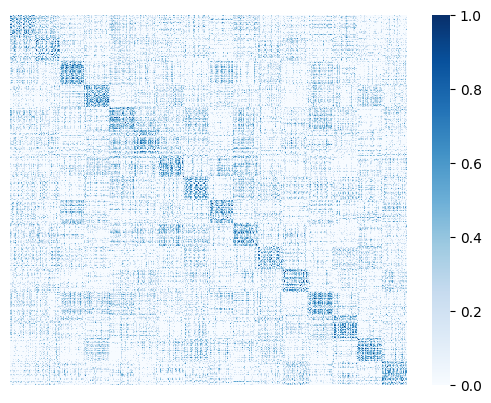}}
\subfigure[Behavior (Variant A)]{
\label{heat4}
\includegraphics[width=4cm,height = 2.8cm]{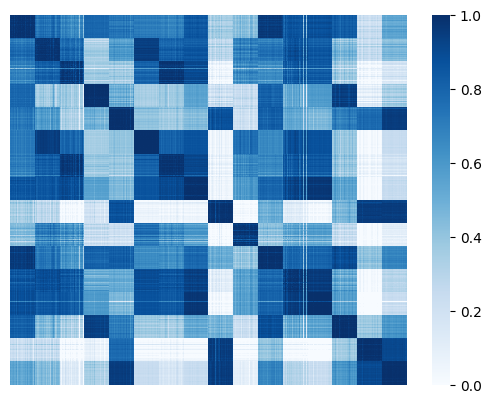}}
\caption{
Independence analysis on ML-100K with predefined 16 intents. Figure(a)-(b) shows the results in IDCL  of behavior and user, respectively, i.e., the cosine similarity between the factors, the diagonal blocks indicates that different factors capture independent information. Figure(c) indicates the result of Variant A (IDCL w/o ICL), the confused high similarity emerge even across different factors. }
\label{corr}
\end{figure*} 

\begin{figure*}[t]
\centering
\subfigure[Intent 1]{
\label{intent2}
\includegraphics[width=4.3cm,height = 3.2cm]{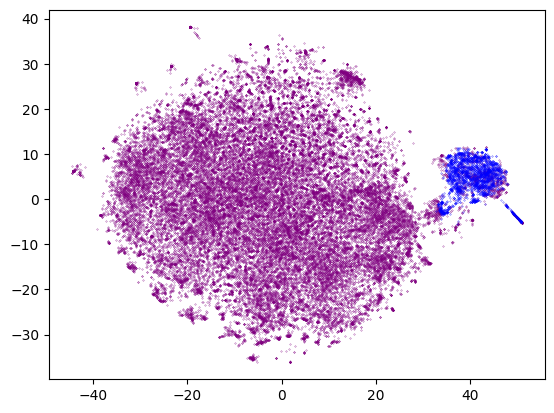}}
\subfigure[Intent 2]{
\label{intent3}
\includegraphics[width=4.3cm,height = 3.2cm]{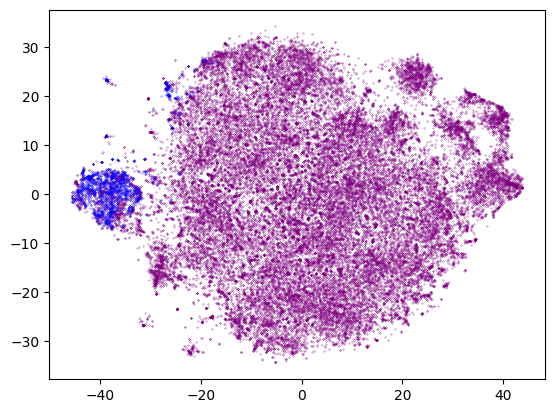}}
\subfigure[Intent 3]{
\label{intent1}
\includegraphics[width=4.3cm,height = 3.2cm]{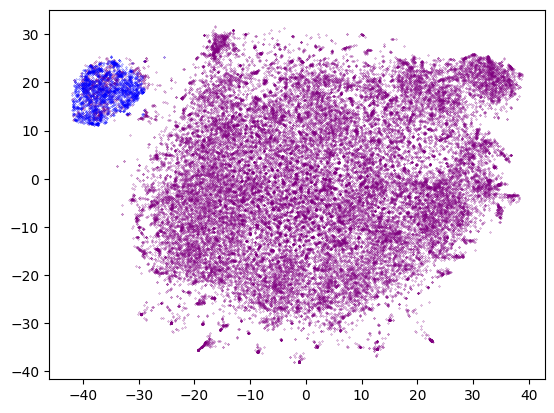}}
\caption{TSNE visualization of the learned intent embeddings on ML-100k. The behavior samples are divided into two disjoint subsets. } 

\label{tsneintent}

\end{figure*}

\subsection{Overall Performance}
 Table \ref{total} summarizes the performance of IDCL and baselines. We have the following observations:
(1) 
Compared with all baselines, the proposed IDCL achieves SOTA performance across all datasets, which demonstrates the effectiveness of our proposed model. This improvement is brought by the informative intent-aware supervision. In particular, it achieves maximum relative improvement over the strongest baseline MacidVAE w.r.t. Recall@20 is $8.52\%$  on ML-100k. IDCL has the most stable performance, i.e. low standard deviation compared to all baselines. Besides, IDCL generally yields more improvement at smaller positions (e.g., top 20 ranks) than at larger positions (e.g., top 100 ranks), indicating that IDCL  promotes to rank related items higher, which is consistent with the requirements of real recommendation scenarios.
(2) 
Among these baselines,  MacidVAE achieves relatively good results in most cases,  which proves the effectiveness of disentangling in recommendation. IDCL surpasses DGCF and MacridVAE across all datasets, confirming the effectiveness of supervision signals in disentangled learning.



\subsection{Independence Analysis}
\paragraph{Are different intents in user behaviors independent of each other?}
Since all behaviors are disentangled into $K$ factors, i.e., $\mathbf{Z}_e=\left[\mathbf{Z}_{e, 1}; \mathbf{Z}_{e, 2}; \ldots; \mathbf{Z}_{e, K}\right]$, each of which indicates a group of identical intents, then  we randomly select $500$ samples from each intent group as $\mathbf{Z}_{e, k}^{*}$,  $k \in [1, K]$.
To investigate if different intents capture mutually exclusive information, we visualize the cosine similarity between each intent in  $\left\{\mathbf{Z}_{e, k}^{*}\right \}_{k=1}^K$.
The result is shown in Figure~\ref{heat1} (the higher similarity corresponds to the darker color),
we can observe that the representations belonging to the same intent (the blocks on the diagonal) are strongly clustered, while the counterparts of different intents are generally independent of each other.
It indicates that  IDCL is able to enforce different intents to be independent, avoiding the information redundancy in behavior representations.


\paragraph{Are IDCL  able to disentangle user representations?} 
Although the disentangling operation in IDCL is only imposed on behaviors, we explore if IDCL also promotes to learn disentangled user representations for deeper analysis.
All user representations are also divided into $K$ parts, 
i.e., $\mathbf{Z}_{u}=\left[\mathbf{Z}_{u, 1}; \mathbf{Z}_{u, 2}; \ldots; \mathbf{Z}_{u, K}\right]$,
each of which indicates a group of identical intents, 
we also randomly selected 500 samples from each intent group as $\mathbf{Z}_{u, k}^{*}$, $k \in [1, K]$.
As in Figure~\ref{heat2}, we visualize the cosine similarity between each intent in  $\left\{\mathbf{Z}_{u, k}^{*}\right \}_{k=1}^K$.
It exhibits the obviously diagonal blocks, i.e., the learned user representations also have a clear disentangled structure despite no explicit disentangling. This indicates that the graph encoder in IDCL is able to separate the distinct, informative intent variations in the interaction graph.



\subsection{Explainability Analysis}
\paragraph{Does the learned representations of intent $k$ capture the semantic of $k^{th}$ intention?}
According to the $k^{th}$ intent representation $\mathbf{Z}_{e, k}$, $k\in[1,K]$, 
we investigate if the sample space is divided into two disjoint subsets on ML-100K, i.e.,  the user behaviors driven by the $k^{th}$ intent and not driven by the $k^{th}$ intent, respectively.
We perform t-SNE visualization~\cite{van2008visualizing} on $\mathbf{Z}_{e, k}$ to analyse the $k^{th}$ intent.
In detail, 
we calculate the distribution of behavior $\mathbf{z}_{e}$ according to Eq.(~\eqref{ek}),  then we color the  points to blue if the confidence of intent $k$ ranks in the top 3, which indicates that behavior $\mathbf{z}_{e}$ is likely driven by the $k^{th}$ intent.
The visualization results of three different intents are shown in Figure~\ref{tsneintent}.
It can be seen that the learned intents can discover behavior partitions of each intent, 
i.e., the clear distance divide between the blue and purple data points.
This indicates that we can characterize whether a user will interact with an item for the reason of intent $k$ just based on the $\Delta d$ dimensional embedding $\mathbf{Z}_{e, k}$, which highlights the significance of disentangling.

\begin{figure}[t]
\centering
\subfigure[Movie]{
    \label{movie}
    \begin{minipage}[t]{2cm}
        \centering
         \includegraphics[width=1.5cm,height = 2.5cm]{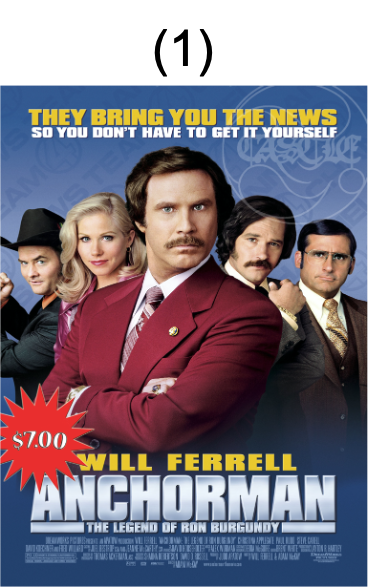}
        \includegraphics[width=1.5cm,height = 2.5cm]{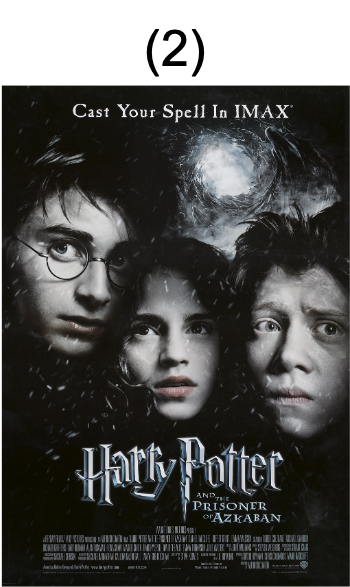}\\
        \includegraphics[width=1.5cm,height = 2.5cm]{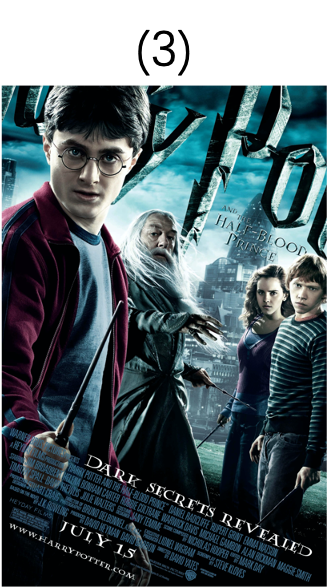}\\
    \end{minipage}%
}
\subfigure[User 1]{
\label{u1}
    \begin{minipage}[t]{3cm}
    
        \centering
        \includegraphics[width=2.5cm,height = 2.5cm]{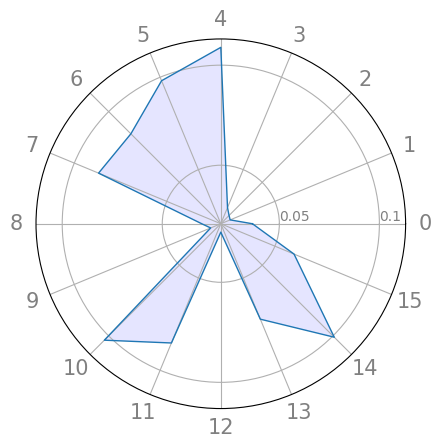}\\
        \includegraphics[width=2.5cm,height = 2.5cm]{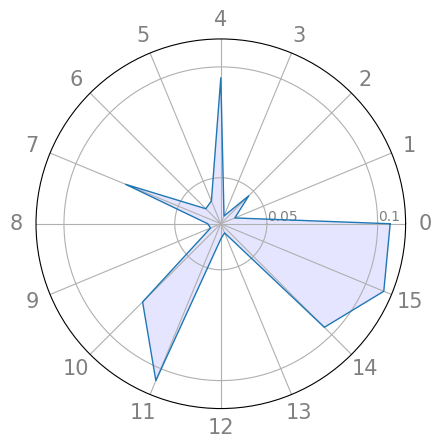}\\
        \includegraphics[width=2.5cm,height = 2.5cm]{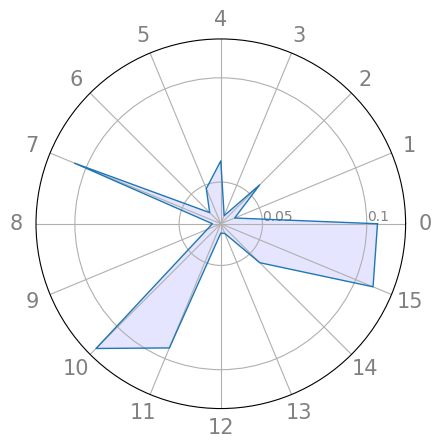}
  
    \end{minipage}%
}
\subfigure[User 2]{
\label{u2}
    \begin{minipage}[t]{3cm}
        \centering
        \includegraphics[width=2.5cm,height = 2.5cm]{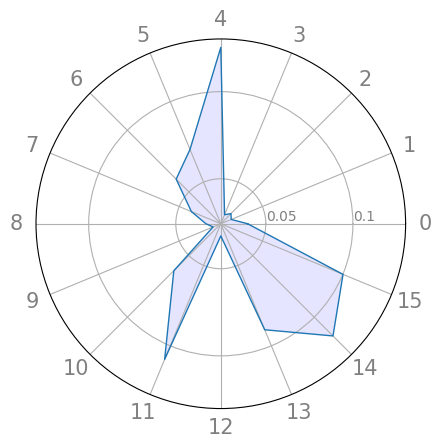}\\
        \includegraphics[width=2.5cm,height = 2.5cm]{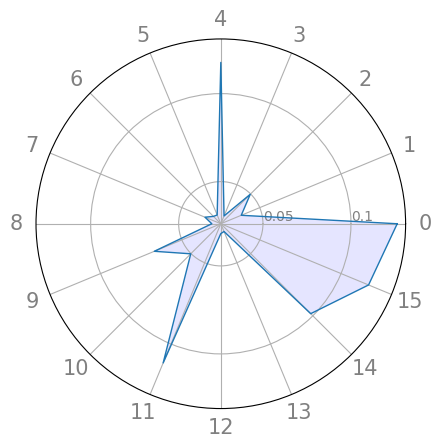}\\
        \includegraphics[width=2.5cm,height = 2.5cm]{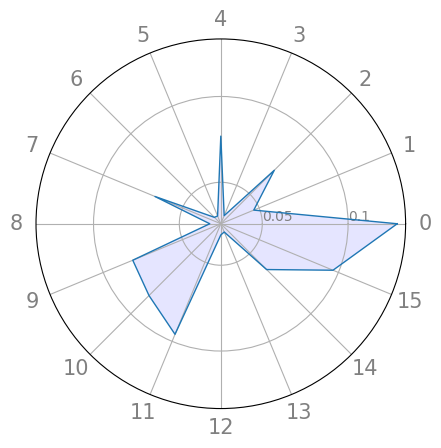}
    \end{minipage}%
}

\centering
\caption{Two users' behavior distributions over all intents for three movies. }
\vspace{-0.2cm}
\label{radar}
\end{figure}

\paragraph{Is the learned user behavior distribution  interpretable?}
We investigate  whether the learned behavior distribution in Eq.~\eqref{ek} can reflect the real reason of why a user interacts with a target item.
In particular, we analyze two users who have both watched the three movies listed in Figure~\ref{movie} on ML-100k, then we visualize their interest distributions over all intents (i.e., $\left\{p_{\theta}(k| e)\right \}_{k=1}^K$ in Eq.~\eqref{ek}) in radar charts as Figure~\ref{u1} and Figure~\ref{u2}, respectively. We disentangle each behavior into $16$ intents,  and the confidence over each intent is the corresponding polar coordinate value.
Then we get some interesting observations.
1) A user watches movies with similar themes 
tends to be inferred similar behavior distributions, i.e., the radar charts of user 1 towards movie 2 and movie 3 (both of Adventure and Fantasy) have very similar shapes. Meanwhile,  it exhibits a different pattern when user 1 interacts with movie 1 (Comedy), e.g., the confidence of intent 5 rises significantly.  We guess that intent 5 likely  indicates ``Comedy" related information.
2) Even if different users watch the same movie, IDCL can still identify different interest distributions reflecting user personality.  i.e., the radar charts of movie 1 from the two users exhibit dissimilar shapes.
This is consistent with the real recommendation scenario.
It indicates that the user's interest distribution is not only related to user personality, but also depends on the characteristics of the target item. 

\begin{table}
\centering

  \small
  \begin{tabular}{lcccc}
    \toprule
        Method & ML-100k &  ML-1M & MtBusiness  \\
        \midrule
        IDCL        & \textbf{0.3235} & \textbf{0.3160}  & \textbf{0.2973} \\
        Variant A & 0.3146            & 0.3122             & 0.2907 \\
        Variant B  & 0.3166            & 0.3146             & 0.2961  \\
        \bottomrule
\end{tabular}

\caption{Ablation studies on the variants of IDCL (Recall@20).}
\label{abl}
\end{table}

\begin{figure}
    \centering
    \includegraphics[width=5.6cm]{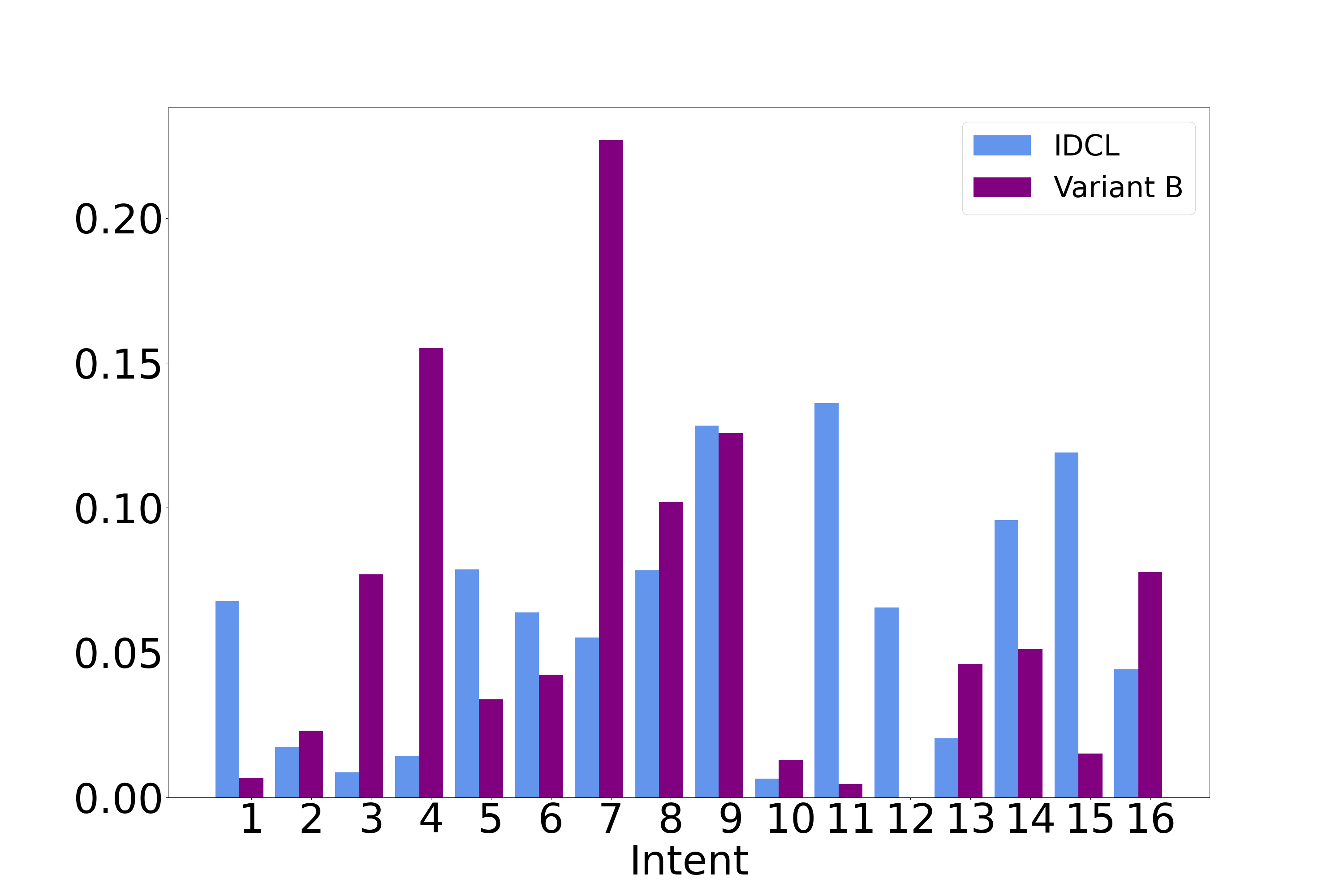}
    \caption{The inferred proportion of behaviors under each intent for IDCL and Variant B, respectively.}
    \label{palar}
\end{figure}

\subsection{Ablation Studies}
To understand the role of each components in IDCL more deeply, we perform ablation studies over two important components, comparing both the recommendation performance and  the quality of the learned representations.
In detail, we design two variants: 
Variant A: IDCL removes the intent-wise contrastive loss (w/o ICL). 
Variant B: IDCL removes the coding rate reduction regularization (w/o CR).

As the results shown in  Table~\ref{abl}, comparing IDCL with Variant A,  we can see that  ICL boosts recommendation performance across all datasets, which indicates that ICL can effectively provide fine-granularity supervised information and assist the representation learning of user and item. 
In addition, we further investigate the impact of ICL on  disentangling in Figure~\ref{heat4}, it exhibits many regions with cluster structure in the off-diagonal blocks, suggesting that features belonging to different intents are highly entangled. It proves that ICL module guarantees the behavior disentangling.

Comparing  IDCL with Variant B in  Table~\ref{abl}, we find that the performance drops slightly  without CR, then we further analyze the impact of CR on representation learning.
To explore if CR can learn discriminative feature to avoid model collapse, i.e., a large percentage of behaviors are assigned to few intents~\cite{XinWang2022DisentangledRL}.
As in Figure~\ref{palar}, we compare the proportion of behaviors under each intent group for IDCL and  Variant B, respectively,
and each behavior is assigned to the intent with the highest probability calculated by Eq.~\eqref{ek}.
We observe that IDCL (blue pillars) distributes the behaviors relatively evenly to each intent.
However, when CR is removed (purple pillars), the behaviors tend to concentrate on few intents, i.e., intent 4 and 7, and even no behavior in intent 12, which weakens the effectiveness of disentangling.
This indicates that CR enhances the dimensional diversity of learned features, which prevents the mode collapse problem.





\section{Conclusion}
In this paper,  we propose IDCL to disentangle user intents and infer behavior distributions. We design a behavior disentangling module to disentangle user intents. We propose a intent-wise contrastive learning module to promote meaningful disentangling and infer the behavior distributions. The coding rate reduction regularization is used to enforce the behaviors of different intents independence. Experimental results show that IDCL substantially improve the  performance and interpretability of recommendation. 
One possible future direction is to incorporate external supervisions to facilitate the disentanglement of interpretable factors. 



\clearpage

\section*{Acknowledgements}
The research was supported in part by the National Natural Science Foundation of China (No. 62172052, U22B2019) and BUPT Excellent Ph.D. Students Foundation (CX2022220).





\bibliographystyle{named}
\bibliography{ijcai23}




\end{document}